# Toward accurate polarization estimation in nanoscopic systems

Sambit Mohapatra[a,*,#], Wolfgang Weber[a], Martin Bowen[a], Samy Boukari[a], Victor Da Costa[a]

[a] Université de Strasbourg, CNRS, Institut de Physique et Chimie des Matériaux de Strasbourg, UMR 7504, F-67000 Strasbourg, France.

Email: tapimohapatra@gmail.com, victor.dacosta@ipcms.unistra.fr

\* Corresponding author

**Abstract:** The nanoscopic characterization of ferroelectric thin films is crucial from their device application point of view. Standard characterization techniques are based on detecting the nanoscopic charge compensation current (switching current) caused by the polarization reversal in the ferroelectric. Owing to various surface and bulk limited mechanisms, leakage currents commonly appear during such measurements, which are frequently subtracted using the device I-V characteristic by employing positive-up-negative-down (PUND) measurement scheme. By performing nanoscopic switching current measurements on a commonly used ferroelectric, $BiFeO_3$, we show that such characterization methods may be prone to large errors in the polarization estimation on ferro-resistive samples, due to current background subtraction issues. Especially, when ferro-resistive behavior is associated with the polarization reversal of the ferroelectric thin film, background current subtraction is not accurate due to the mismatch of the I-V characteristics for the two polarization states. We show instead that removing the background current by an asymmetric least squares subtraction method, though not perfect, gives a much better estimation of the ferroelectric properties of the sample under study.

Ferroelectric materials find applications in numerous fields ranging from energy harvesting, sensors and transducers, and heat transfer applications. A large number of electronic devices based on ferroelectrics such as random-access memory (FeRAM),[1,2] field effect transistors (FeFET),[3] tunnel junctions (FTJ),[4,5] photovoltaics,[6] optoelectronic devices,[7] resistive switches,[8] and ferroelectric memristors[9] have been explored. Further, the possibility to control the polarization state using an electric field makes ferroelectrics suitable for energy efficient nanoelectronics and spintronic devices.[10] More recently, ferroelectric negative capacitance,[11] and domain wall memory[12,13] devices are adding to the functionality and applicability of ferroelectric materials in device technology. The continuously growing applications of ferroelectric materials in the world of nano devices underscore the need to characterize and understand the ferroelectric properties at the nanoscale.

Standard techniques for quantitative characterization of ferroelectric materials involve the detection of polarization switching currents. A ferroelectric material is characterized by measuring

---

[#] Present address: Centre for Nanoscience and Nanotechnology (C2N), University of Paris-Saclay

the polarization-versus-voltage hysteresis loop in a capacitor geometry with the ferroelectric as dielectric. The polarization is estimated as the time integration of the measured compensation current (i.e. polarization charges) per unit area. The polarization charge compensation current, commonly called the polarization switching current, that appears in these experiments due to polarization switching is purely macroscopic in nature as the lateral dimensions of the capacitors are usually at least several 10s of micrometers.

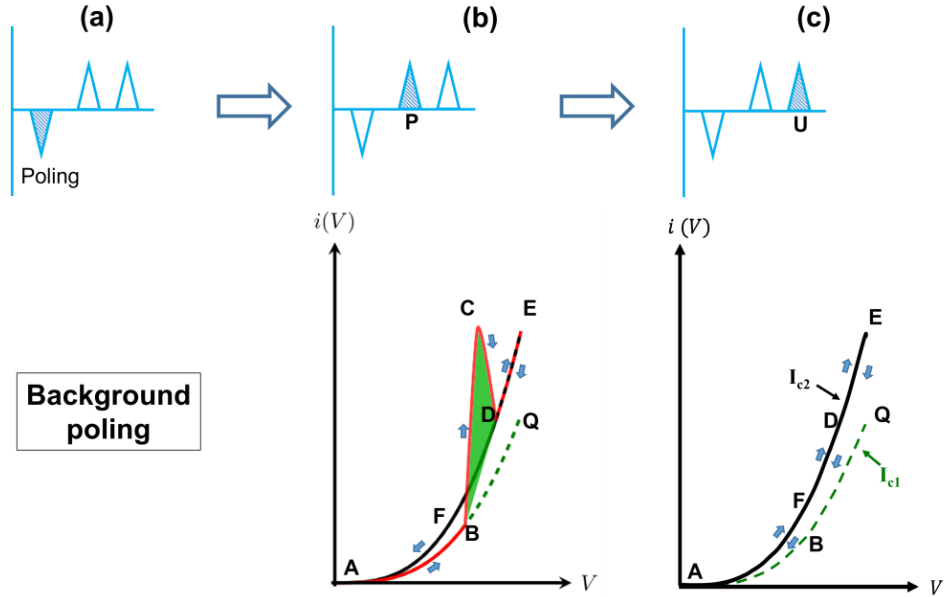

**Figure 1.** Schematics showing the voltage pulse sequence for the PUND technique (only the PU part) for the case when I-V curve switches from high to low resistance state. The triangular pulses represent the applied voltage scheme. The consequence of each voltage pulse, as represented by the shaded pulse, is shown below the respective voltage schematic. The first pulse, background poling (a) (negative) is used to polarize the nanoscopic region or a larger region surrounding the nanoscopic one along a certain vertical direction (upward). The second pulse, (b) results in the sum of both the conduction current ($I_C$) and the ferroelectric polarization switching current ($I_F$), $I_P = I_F + I_{C1}$. The ideal I-V curves corresponding to the two polarization states are represented by black (AFDE) and green curve (ABQ), with the initial polarization state corresponding to the green curve ABQ. The solid red curve (ABCDE) represents the actual curve ($I_P$) corresponding to the applied voltage pulse, showing the transition from the green to the black curve. The small blue arrows show the actual direction of current evolution. The green shaded region represents an exaggerated view of the ferroelectric polarization switching current peak. The portion BQ on the green curve remains inaccessible during this pulse. The third pulse, (c) is similar to the second one, however, it results in only the background conduction leakage current ($I_U = I_{C2}$) as indicated by the solid black curve (AFDE). The entire curve corresponding to the opposite polarization state, shown in green dashed lines (ABQ), remains inaccessible during this pulse.

Alternatively, to gain insights into the properties at the nanoscale, there have been recent attempts to utilize conductive probe tips, for example in a standard Piezoresponse Force Microscopy (PFM) setup, to detect the local polarization switching[14,15] either from uncapped or capped ferroelectric thin films. Such measurements are further supplemented by measuring the simultaneous strain response of the nanoscopic region during polarization reversal, where strain response refers to the electromechanical response of the region which is in contact with the probe via inverse piezoelectric effect.[14]

However, whether at the macro- or nanoscale, the major obstacle that lies at the center of such measurements is the possibility to encounter significantly large currents originating from sources other than the polarization switching.[16]

One such prominent additional current is the conduction current, the presence of which is a well-known issue when it comes to the measurement of polarization switching currents in macroscopic ferroelectric capacitor devices. Owing to either quantum tunneling leakage, various electrode-dielectric interface limited mechanisms and dielectric bulk limited mechanisms or a combination of these, conduction currents can flow through a ferroelectric capacitor.[17–21] This usually results in the presence of a non-linearly increasing background current during an I-V measurement. Such a current background may stand in the way of an accurate observation of the polarization switching current peak, particularly in the case of a nanoscopic devices where the peak height of the switching current can be overwhelmed by the magnitude of the conduction current.[22] Furthermore, the presence of a conduction current leads, after the current-time integration process, to an erroneous estimation of the polarization, as it prevails over the switching current, thereby inhibiting an accurate quantitative ferroelectric characterization.

Conventionally, the problem of leakage currents in macroscopic capacitors is mitigated by employing the Positive Up-Negative Down (PUND) technique,[23] where the device is subjected to a negative voltage pulse followed by two consecutive positive voltage pulses (fig. 1a,b,c). With the bottom electrode grounded, the negative pulse sets the sample in an up polarization. Then the first positive pulse causes the actual polarization reversal and results in a net current that is the sum of the conduction current ($I_C$) and the polarization switching current ($I_F$). If the remanence of the polarization is 100%, the second positive pulse results only in the conduction current. Thus, the difference between these two measured current values removes the large conduction current and the smaller polarization switching current peak is retrieved. The time integration of this I-V curve thus obtained divided by the area of the switched region, as obtained by PFM domain mapping, generates the polarization hysteresis loop. The application of this technique has been extended recently to nanoscopic measurements.[15] Other techniques are used to correct for the leakage current without the need to apply any peculiar pulses to prepare the sample (see e.g. reference 24), but they consider that the leakage current has an ohmic behavior.

These procedures to remove the conduction current are inefficient if the conductance depends on the polarization state of the ferroelectric. Yet, having a conductance that depends on the polarization state is common, for example, in ferroelectric tunnel junctions.[25] In such cases, the polarization-dependent height of the tunnel barrier translates into different resistance states. Taking advantage of this polarization dependence, non-destructive readouts in ferroelectric memories have been proposed recently.[5,26] Similarly, ferro-resistive systems, where the reversal in ferroelectric polarization is accompanied by a change in the device resistance, are proving essential for future electronic and computing applications.[8,27]

Starting with the initial polarization state as up or down, the device presents two I-V characteristics corresponding to the two polarization states of the ferroelectric spacer (figure 1), as long as the voltage doesn't modify the initial polarization. Once the voltage is sufficient to initiate a reversal, the device transits from one I-V characteristic to the other (from green (ABQ) to black (AFDE) curve in figure 1b). When the reversal is complete, the device resistance is given by the I-V characteristic corresponding to a state with a polarization opposite to the initial one (figure 1c). The polarization reversal is accompanied by the appearance of ferroelectric polarization switching current (green shaded region in figure 1b). Thus, during the reversal, the switching current peaks are superimposed on the conduction current in the I-V curve. We see that if the I-V characteristics do not depend on the polarization, the black and green curves will be superimposed and the PUND procedure will compensate well for the conduction current, thereby resulting in an accurate estimation of the ferroelectric polarization switching current, and hence, of the polarization magnitude. However, if the I-V characteristic depends on the polarization, the PUND technique may be inefficient to compensate for the conduction current appearing during the transition from one characteristic to the other, as the I-V curves corresponding to two polarization states do not coincide (figure 1b and 1c).

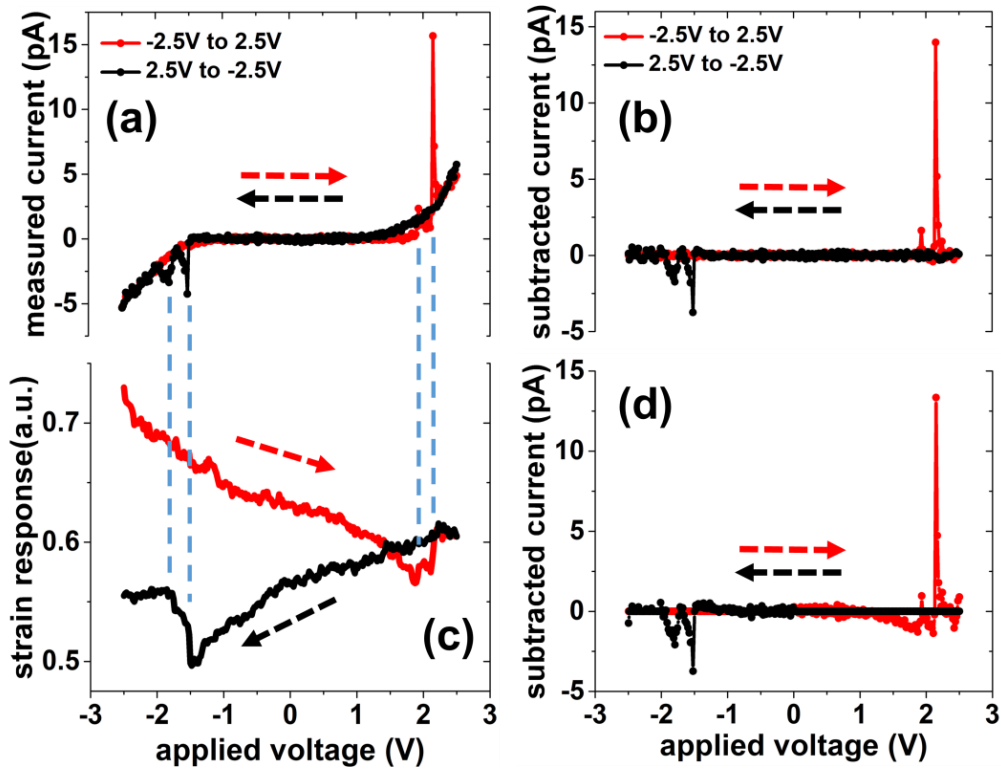

**Figure 2.** I-V measurements on a 65 nm uncapped BFO film with Pt probe tip. (a) shows the I-V measurements without employing PUND technique. (b) shows the approximate background-subtracted I-V curves as obtained in (a). The background subtraction is carried out by asymmetric least squares method. (c) shows the piezoelectric strain response (raw data) during the I-V measurement. The synchronous jumps in this curve are characteristics of actual polarization switching occurring in the nanoscopic region under the tip. (d) shows the I-V curves obtained with the PUND technique. The red and black curves represent the two directions of applied voltage, from -2.5 V to 2.5 V and 2.5 V to -2.5 V, respectively.

As shown in figure 1b (second pulse), the solid red curve consists of the ferroelectric polarization switching current ($I_F$) peak superimposed on the conductance current ($I_{C1}$) for the particular polarization state ($I_P = I_{C1} + I_F$). In the next consecutive pulse (third pulse), the measured current $I_U = I_{C2}$, as no switching takes place for this pulse. Conventionally, the ferroelectric polarization switching current is estimated as $I_F = I_P - I_U$. However, due to the ferro-resistive behavior, the conductance current ($I_U = I_{C2}$) appearing (figure 1c) is the current in the opposite polarization state of the ferroelectric and is different from that in the previous pulse (figure 1b; $I_{C1}$). Thus, it is not possible to obtain an accurate value for $I_F$ by subtracting the conductance current $I_{C2}$ from $I_P$. The error introduced by such a subtraction will depend on the difference between $I_{C1}$ and $I_{C2}$, which stems from the difference in the I-V characteristics corresponding to the two polarization states. Supplementary figure S1 shows the schematics similar to figure 1 for the scenario corresponding to the case when the I-V curve switches from a low to high resistance state during the polarization reversal.

In this article, by considering the case of $BiFeO_3$ (BFO), a well-studied ferroelectric, on a standard substrate we show how the above discussed issue leads to difficulties in accurate nanoscopic ferroelectric characterization, when the conductance depends on the polarization state and propose a technique to resolve it. As our system under study, a BFO sample of thickness 65 nm was grown by pulsed laser deposition on 70 nm of $La_{0.7}Sr_{0.3}MnO_3$ (LSMO) supported by a $SrTiO_3$ (STO) single crystal substrate.[28] Such systems are commonly studied in connection to their magnetoelectric applications in spintronics.[28] The sample was exposed to atmosphere before being transferred to an ultra-high vacuum environment for the ferroelectric characterizations that were carried out using a standard PFM setup (Scienta Omicron UHV VT-SPM). To obtain the I-V characteristics of the BFO film, current is measured via a Pt probe tip in contact with the surface of the film during the application of an electric voltage, which was varied as a linear ramp from -2.5 V to +2.5 V (forward) and then from +2.5 V to -2.5 V (backward) at a cycle frequency of approximately 2 Hz.

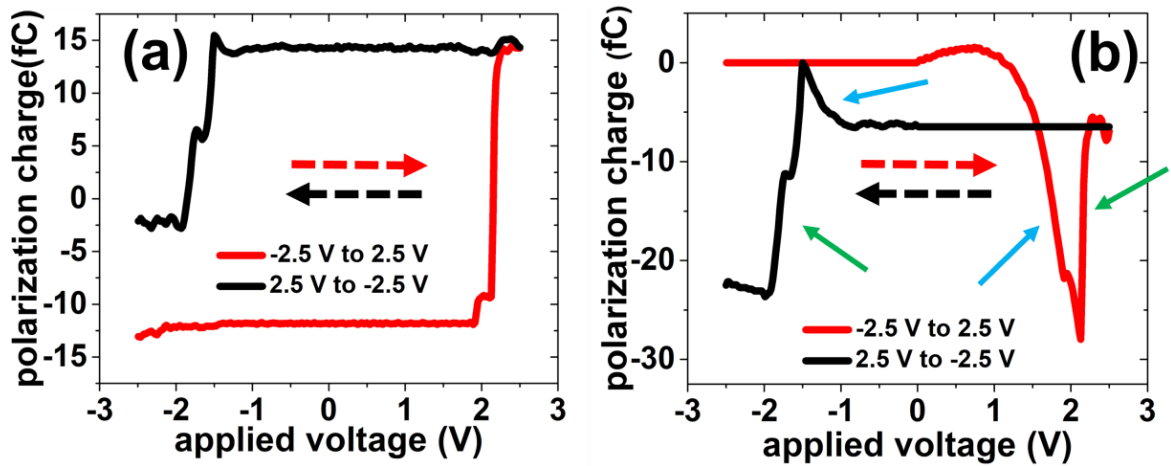

**Figure 3.** Polarization hysteresis loops obtained (a) by the integration of 2b and (b) of 2d. The green and blue solid arrows in (b) indicate the change in polarization charge due to actual switching events and due to the error in the background subtraction using PUND technique, respectively. The dashed arrows indicate the direction of the applied voltage. The curve (a) was shifted so that it is vertically symmetric at V=0.

We observe multiple switching current peaks (figure 2a) that are superimposed on an s-shaped background (see figure S3 for the background curve). The fact that the current peaks correspond to a stepwise reversal of the polarization is supported by simultaneous multiple jumps in the piezoelectric strain response curves as shown in figure 2c (blue dashed lines). Starting from -2.5 V, the ferro-resistive behavior is clear from the two distinct I-V curves in figure 2a before (red curve from -2.5 V to ~+1.8 V) and after the polarization reversal (black curve from +2.5 V to ~ -1.5 V) (see an expanded view of fig 2a in the supplementary figure S2).

Next, to estimate the polarization, we try to remove the background due to the conduction current by employing two methods: (i) fitting the background of the I-V curve by an asymmetric least squares approximation method and (ii) a background subtraction combining the forward and backward curves in figure 2a (red and black curve respectively) as done in the PUND technique.

In the first method, we approximate the I-V curve background using an asymmetric least squares method[29,30,31] and subtract the as-obtained background from the I-V curve to isolate the current due only to the polarization reversal. The method combines a smoother with asymmetric weighting of deviations from the smooth trend to get an effective baseline estimator. No prior information about peak shapes or baseline is needed. It is thus well suited to subtract the background from I-V curve measured on ferroelectrics, where the background has multifactorial origins and is difficult to model. We used the asymmetric least squares fitting algorithm provided by the commercial software Origin from the OriginLab Corporation. The baseline subtracted I-V curves are shown in figure 2b (See supplementary figure S3 for the baseline).

In the second method, to mimic the PUND technique, we subtract in figure 2a, the black curve from the red one for positive biases and the red curve from the black one for negative biases to obtain the curve shown in figure 2d. The black and red curves represent the U and P pulses for positive biases and N and D pulses for the negative biases, respectively. Also, when the PU and ND curves are combined to generate figure 2d, currents for the negative biases for the PU part and positive biases for the ND part (no polarization switching region) are kept as zero. Ideally, the U and D pulses should be separate ramps from 0 to 2.5 V and 0 to -2.5 V, respectively. However, in our case, their directions are reversed. Nevertheless, we have verified that this does not affect the measured current curves as the conduction current is independent of applied bias direction if polarization switching does not take place.

The subtracted current is negligibly small between ~-1.5 V and ~+1.8 V in figure 2b, in the bias range where there is no evidence of reversal. The conduction current is thus well subtracted in this region. However, in figure 2d there is still some current in the same region. Especially, the negative currents that appear while sweeping the voltage from -2.5 V to +2.5 V (red curve in figure 2d) are not physical; their appearance is merely an inevitable consequence of a poor background subtraction. This is inevitable as $I_{C2}$ is higher than $I_P$ for a major section of the applied voltage on the positive side. Evidently, the PUND technique is unable to entirely remove the conduction current even when there is no reversal, precisely because the conduction characteristics depend on the polarization state.

The polarization hysteresis loops (polarization charge versus applied voltage) shown in figure 3a and 3b are obtained by the integration of the curves in figure 2b and 2d, respectively. Note that polarization is related to the polarization charge divided by the area of the switched region,[14,15] which could not be captured in our experiment. However, is has no impact on the analysis as the polarization charge is directly proportional to the switched area for a fixed polarization. It is clear from figure 3b that the PUND-based results suffer from a severe distortion of the hysteresis loop, which makes an estimation of the polarization very difficult. Evidently, this is due to the incomplete conduction current elimination with the subtraction mimicking the PUND technique (figure 2d). On the contrary, figure 3a presents a less distorted hysteresis loop, but the loop is not closed. This means that there is still an error in the polarization determination. This error can have two different origins.

As we have shown previously,[14] a loss of polarization can be observed because some ferroelectric switching current can be buried within the instrumental noise, especially at the nanoscopic scale. A second possibility, cumulative with the first one, is that the conductance depends, as here, on the polarization state. The current that is measured during the reversal has a contribution from the change in polarization (switching current) plus one due to an evolution of the conductance. Therefore, a correct understanding of the mechanism of current conduction through the ferroelectric and its dependence on the polarization state, and more importantly, on the kinetics of the polarization reversal process is essential to identify this contribution. To our knowledge, there is no straightforward way of perfectly separating one contribution from the other. The asymmetric least squares method, although not perfect, seems to be the best available alternative so far in terms of producing an approximate but acceptable P-V hysteresis loop.

To conclude, we show that the nanoscopic characterization of polarization in ferroelectric thin films may not be straightforward when the current-voltage characteristic is dependent on the polarization state. The difference in I-V characteristics in both the polarization states of the ferroelectric leads to a background that cannot be accessed and subtracted in the commonly used PUND technique. Further, as the state of the polarization evolves it is difficult to approximate the current background during a switching event. This can lead to an erroneous detection of the switching current transient, thereby hindering the study of fundamental properties of the ferroelectric. In the absence of any standard technique to isolate the characteristic background from the switching current peaks, asymmetric least squares background subtraction methods may be employed as an alternative to improve accuracy when characterizing polarization hysteresis loops and ferroelectric properties.

Acknowledgements:

We would like to thank the Unité Mixte de Physique CNRS/Thales lab for the sample preparation. We would like to acknowledge the agencies ANR-DFG for funding support on the project ORINSPIN (ANR-16-CE92-005-01).